# MATERIAL CHEMICAL COMPOSITON IMPACTS ON THE BAND ALIGNMENTS: PRELIMINARY RESULTS ON THE 3D/2D PEROVSKITE INTERFACES

Philippe Baranek[1, 2, •]
[1] EDF R&D, Department SYSTEME, EDF Lab Paris-Saclay, 7 bd Gaspard Monge, F-91120 Palaiseau, France
[2] IPVF, Institut Photovoltaïque d'Ile-de-France, 18 boulevard Thomas Gobert, F-91120 Palaiseau, France

ABSTRACT: While hybrid organic-inorganic halide perovskite solar cells have achieved remarkable certified efficiencies, their widespread industrial and societal adoption is hindered by instabilities against light, heat, and moisture. A strategy to mitigate issues, such as moisture sensitivity, involves depositing 2D perovskites on 3D thin films for passivation, but their precise impact on cell performance requires detailed understanding. This work employs atomistic first-principles simulation, based on hybrid functionals, to investigate the structural and electronic properties of 2D/3D interfaces. Specifically, it focuses on interfaces formed by $(PEA)_2PbI_4$ (PEA = $C_6H_5CH_2CH_2NH_3$, as 2D perovskite) and complex 3D perovskites ($Cs_{0.125}MA_{0.14}FA_{0.735}Pb(I_{0.87}Br_{0.13})_3$, with MA = $CH_3NH_3$ and FA = $CH(NH_2)_2$) which are ones of the most used perovskites. The approach is based on optimized hybrid functional methods to consistently determine work function (WF), electron affinity ($\chi$), and band offset ($E_{BO}$) across different layers. Preliminary results demonstrate that WF are strongly dependent on the cleavage and chemical nature of the surfaces. Furthermore, the creation of interfaces significantly impacts macroscopic potential, with a noticeable slope linked to the polarization of the 3D materials due to ionic substitutions. The band alignment is also highly sensitive to the termination and chemical nature of the perovskites, with valence and conduction $E_{BO}$s varying from 0.2 to 0.7 eV and of approximately 0.25 eV, respectively. This ongoing work aims to provide a systematic description of these effects, with further investigation into the impact of dimensionality loss on perovskite stability.

## 1 INTRODUCTION

Hybrid organic-inorganic halide perovskite solar cells have emerged as a revolutionary technology in photovoltaics, achieving power conversion efficiencies that rival traditional silicon-based devices [1, 2]. Despite their rapid advancements and record efficiencies, their long-term stability under operational conditions (light, heat, moisture) remains a significant challenge for commercialization. To address these instability issues, various strategies are being explored, including the incorporation of 2D perovskites as passivation layers on top of 3D perovskite thin films [3]. However, the intricate interplay between the chemical nature, dimensionality, and interfacial properties of these materials makes it challenging to predict their precise impact on solar cell performance a priori.

In fact, the chemical compositions and dimensionality (3D (bulk) and 2D (surfaces, interfaces, thin films)) of perovskites strongly influence the performances of solar cells. Their impact concerns mainly the electronic properties, the domains and surfaces stabilities of the different compounds.

Understanding the electronic properties at these interfaces, particularly band alignment, is crucial for optimizing charge transport and minimizing recombination losses. The primary goal of this work to evaluate the influence of the surfaces and interfaces formation on the bulk properties of the absorbers. It presents the application of atomic theoretical materials modelling, based on density functional theory (DFT) and Hartree-Fock (HF) approximation, to study the structural, electronic, properties of the 2D/3D interfaces constituted by the $(PEA)_2PbI_4$ (2D perovskite) and the complex perovskites ($Cs_{0.125}MA_{0.14}FA_{0.735}Pb(I_{0.87}Br_{0.13})_3$, 3D perovskite) two most commonly used perovskites for the photovoltaic applications. This choice is also motivated by its novelty to the best of my knowledge, there isn't study on this kind of complex interface, and it represents the opportunity to test the reliability of the different approaches.

It is based on hybrid functional approaches which permits to define the Hamiltonian best suited for describing the properties of a given family of materials. It has been used to efficiently study the impact of perovskite phase transition on the cell performance [4, 5] or the optimization of the chemical composition of inorganic perovskites [6]. Hybrid functional approach also allows to obtain a homogeneous description (i.e. at the same level of precision for various data electronic affinity, band gap, and band offset) of the different layers.

The determination of band offsets and alignments at interfaces is a complex, ongoing aspect of this study. Some of the methods employed in this work have been applied to the determination of the band offset at the different interfaces or to study the effect of adsorbates on the electronic properties of simple semiconductors [7 – 11]. In this work, the proposed method is based on the combination of supercell approach, slabs model and 3D description of the interfaces.

This approach is based on a crystallographic description of the complex 3D and 2D perovskites which will serve as basic atomic structure to create the different surfaces and interfaces. We first describe this crystallographic model. At the first principles level, it is used to evaluate the hybrid exchange-correlation functional adapted to these two types of perovskites. The performances of this functional on the structural and electronic (band gap, valence band edge and electron affinity) properties are then systematically investigated on the simple $FAPbX_3$ (X = Br and I), and the complex 2D and 3D perovskites. Next, the impacts of the surfaces formation on these properties are then evaluated on the isolated surfaces. Finally, the preliminary results on the band offsets and alignments through various interfaces are then described focusing on the influence of the interface chemistry on these data and the computational difficulty to treat these kinds of complex systems.

•Email : philippe.baranek@edf.fr

## 2 METHODOLOGICAL ASPECTS

For the complex perovskites, we define a crystallographic structure allowing the ordering of the different chemical compounds inside the lattice and through the different interfaces (see figure 1).

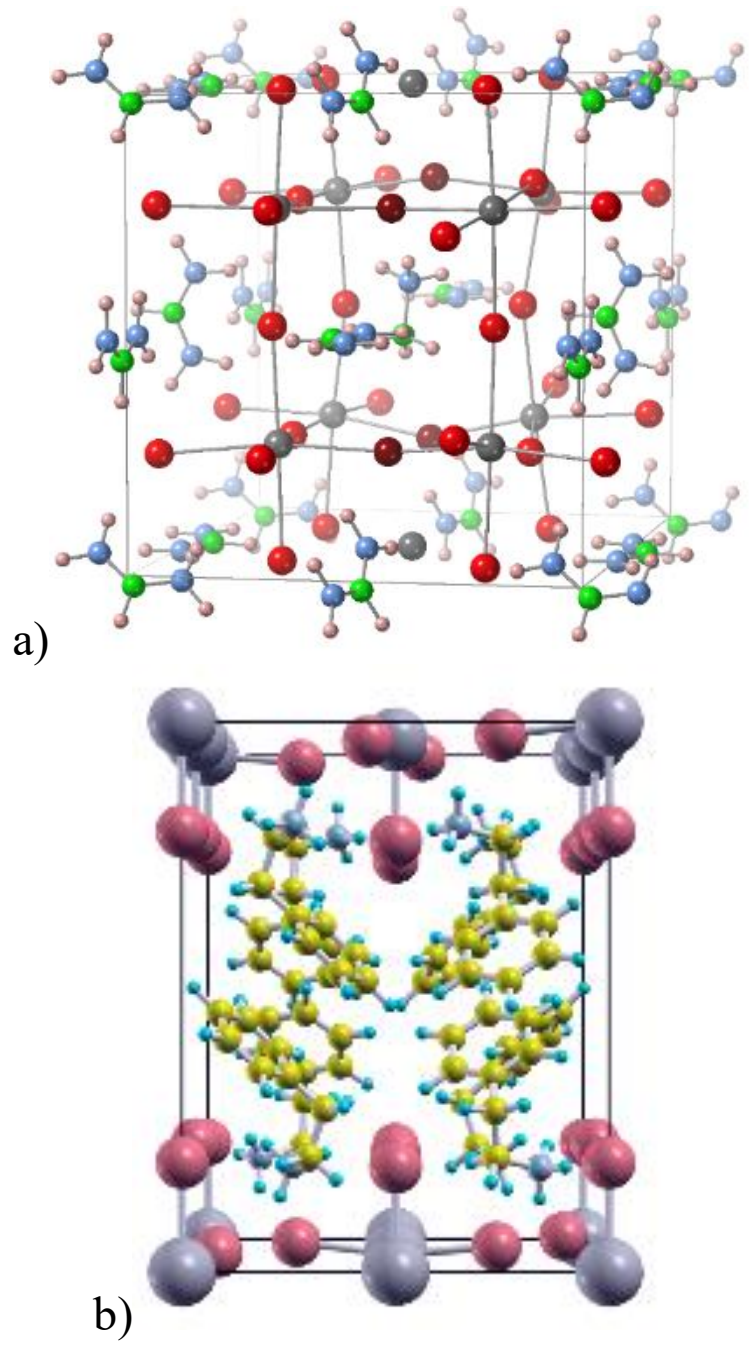

**Figure 1:** Crystal models for a) the 3D perovskite $Cs_{0.125}MA_{0.14}FA_{0.735}Pb(I_{0.87}Br_{0.13})_3$ and b) for the 2D perovskite $(PEA)_2PbI_4$ (tetragonal phase, space group *P4*).

For the 3D perovskite, the crystallographic structure is based on the $FAPbX_3$ crystal description defined in Ref., and the optimization of the different substitutions follows the procedure defined in Ref. [12].

First-principles calculations have been performed with the use of the CRYSTAL code [13 – 15]. This program enables to solve both the Hartree–Fock (HF) and the Kohn–Sham (KS) systems of equations, combining them within a hybrid scheme. In this work, the Hamiltonian combines 5.625% of HF exact exchange with the PBE exchange correlation functional [16]. For the $FAPbX_3$ perovskites, benchmark calculations have been detailed in [17]: the average errors with respect to the experimental data dispersion in the values of the lattice parameters and band gaps are 2% (± 1%) and 3% (± 2%), respectively.

At the first-principles level, determining the work function (WF) and electron affinity (EA) necessitates modeling various surfaces to establish the vacuum potential. In this framework, their definitions are:

$$WF = E_{vac} - E_{Fermi} \quad (1)$$

$$EA = E_{vac} - E_{CBM} \quad (2)$$

where $E_{vac}$, $E_{Fermi}$, and $E_{CBM}$ represent the vacuum potential, the Fermi level energy, and the energy of the lowest conduction band, respectively.

These different data have been evaluated combining the calculation of the valence band edge energy based on the Janak theorem and of the void energy via the determination of the macroscopic potential through the (001) surface of different materials considered in this work. With CRYSTAL, surfaces properties can be studied within different slab models [13 – 15]. In this paper, the two-dimensional (2D) slab model is applied [13 – 15]: Surfaces are modeled as slabs of finite thickness, periodically repeated in two directions (x, y) but not in the third dimension (z), which corresponds to the vacuum. The slab thickness must be sufficient to ensure that the Fermi energy is not influenced by surface states and that the average macroscopic potential remains constant throughout the surface (as illustrated by figure 2).

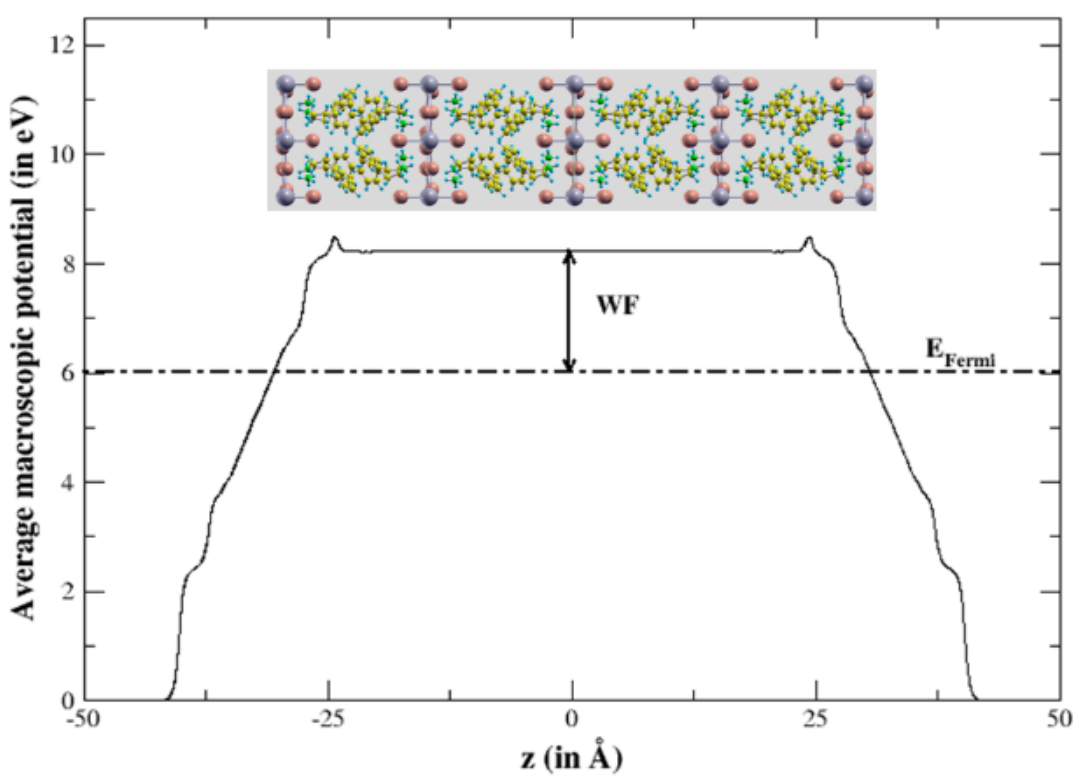

**Figure 2:** Average macroscopic potential determined with the slab model of the (001) surface of the 2D perovskite $(PEA)_2PbI_4$; The plateau gives the vacuum level. The $E_{Fermi}$ position is arbitrary to illustrate the method.

For the determination of band offsets at interfaces, the approach relies on combining results from two isolated surfaces with an interface model. The isolated surfaces allow for positioning the valence band maximum ($E_{VBM}$) relative to the average electrostatic potential across the surface. The interface model then determines the lineup of the average electrostatic potential across the interface, $\Delta V$. The band offset $E_{BO}$ is thus calculated as:

$$E_{BO} = E_{VBM\text{-}S1} - E_{VBM\text{-}S2} + \Delta V \quad (3)$$

Interface modeling presents significant computational challenges, as both surfaces must be thick enough to achieve converged electronic properties, and the interface area must be sufficiently large to accommodate any mismatch between materials. The calculations are realized on converged cells of 2D/3D interfaces: The mismatch between materials is less than 1 % ($a_{3D}$ and $a_{2D}$ are 6.323 and 6.247 Å, respectively) facilitating the modeling. The obtained band gaps of the 2D and 3D perovskites are 2.55 and 1.96 eV, respectively, coherent with experiments [18]. These systems can reach up to 1650 atoms, with slab thicknesses up to 80 Å for the 2D and 3D perovskite surfaces.

## 3 RESULTS AND DISCUSSION

The Figure 1 and Table I give unit cells used to realize the calculations on the 2D and 3D perovskites and the results obtained on their lattice parameters, band gaps, valence band edges and electron affinities: The 3D pristine cell is 89-atoms cell which enables to begin to consider the influence of the distribution of the molecular entity across the lattice on the structural, vibrational and optoelectronic properties of this material; the 2D cell description is based on its high symmetry tetragonal phase. As noticed in previous publications, these units combined with the optimized Hamiltonian to reproduce the properties of the perovskites allows to obtain an estimation of the lattice parameters and band gaps with an average error of 2. and 5 %, respectively, with respect to the available experimental data as benchmark for various types of perovskites [4 – 6, 17, 19, 20].

**Table I:** Calculated bulk properties of examples of perovskites treated in this work. Lattice parameters (a and c in Å), band gap ($E_g$ in eV), valence band edge ($E_{VBM}$ in eV) and electron affinity ($\chi$ in eV). Experimental data available are given for comparison.

| Material | a | c | $E_g$ | $E_{VBM}$ | $\chi$ |
|---|---|---|---|---|---|
| $MAPbI_3$ [4, 5] | | | | | |
| Calc. | 6.368 | | 1.68 | 5.47 | 3.79 |
| Exp. | 6.329, 6.308 | | 1.50 – 1.69 | 5.05 – 5.92 | 3.45 – 4.10 |
| $FAPbBr_3$ [17] | | | | | |
| Calc. | 8.655 | 12.067 | 2.18 | 5.73 | 3.55 |
| Exp. | 8.415 – 8.438 | 11.892 – 11.926 | 2.15 – 2.28 | | |
| $FAPbI_3$ This work | | | | | |
| Calc. | 8.659 | 12.073 | 1.73 | 5.45 | 3.72 |
| 2D This work | | | | | |
| Calc. | 6.247 | 16.207 | 2.55 | 5.59 | 3.04 |
| 3D This work | | | | | |
| Calc. | 6.323 | | 1.96 | 5.89 | 3.94 |

Table II summarizes the calculated band gaps (bulk and (001) surfaces), work functions, and electron affinities for $MAPbI_3$, and $(PEA)_2PbI_4$ for different terminations of the (001) surfaces.

**Table II:** Calculated bulk and (001) – surfaces electronic properties of for $MAPbI_3$ and $(PEA)_2PbI_4$. Band gap ($E_{g_bulk}$ and $E_{g_surf}$ in eV), valence band edge ($E_{VBM}$ in eV) work function (WF in eV), electron affinity ($\chi$ in eV), and effective electron affinity of the surfaces ($\chi_{surf} = WF – E_{g\text{-}surf}$ in eV).

| | $MAPbI_3$ | | $(PEA)_2PbI_4$ | |
|---|---|---|---|---|
| $E_{g\text{-}bulk}$ | 1.68 | | 2.55 | |
| $E_{VBM}$ | 5.47 | | 5.59 | |
| $\chi$ | 3.79 | | 3.04 | |
| | Surface's terminations | | | |
| | $PbI_2$ | MAI | $PbI_2$ | PEAI |
| $E_{g\text{-}surf}$ | 1.65 | 1.81 | 1.90 | 1.13 |
| WF | 5.19 | 5.33 | 5.49 | 4.39 |
| $\chi_{surf}$ | 3.54 | 3.52 | 3.59 | 3.26 |

At this stage, the impact of the surface's terminations on their relative chemical and thermodynamical stability is not covered. It aims to illustrate that the chemical nature of the surface profoundly impacts the obtained data. For (001)-$MAPbI_3$ MAI terminated surface, an increase in the band gap is observed, which correlates with a more ionic character of this slab. Conversely, for both terminations of the (001)-2D surface, a decrease in the band gap is linked to a more ionico-covalent nature of these different slabs. These differences in chemical bonding and surface termination also account for the variations in band gap work function and effective electron affinities of the surfaces between the two types of terminations for each perovskite. It illustrates that the lowering of the system's dimensionality associated with the surface cleavage can strongly influence the electronic properties of the isolated surfaces.

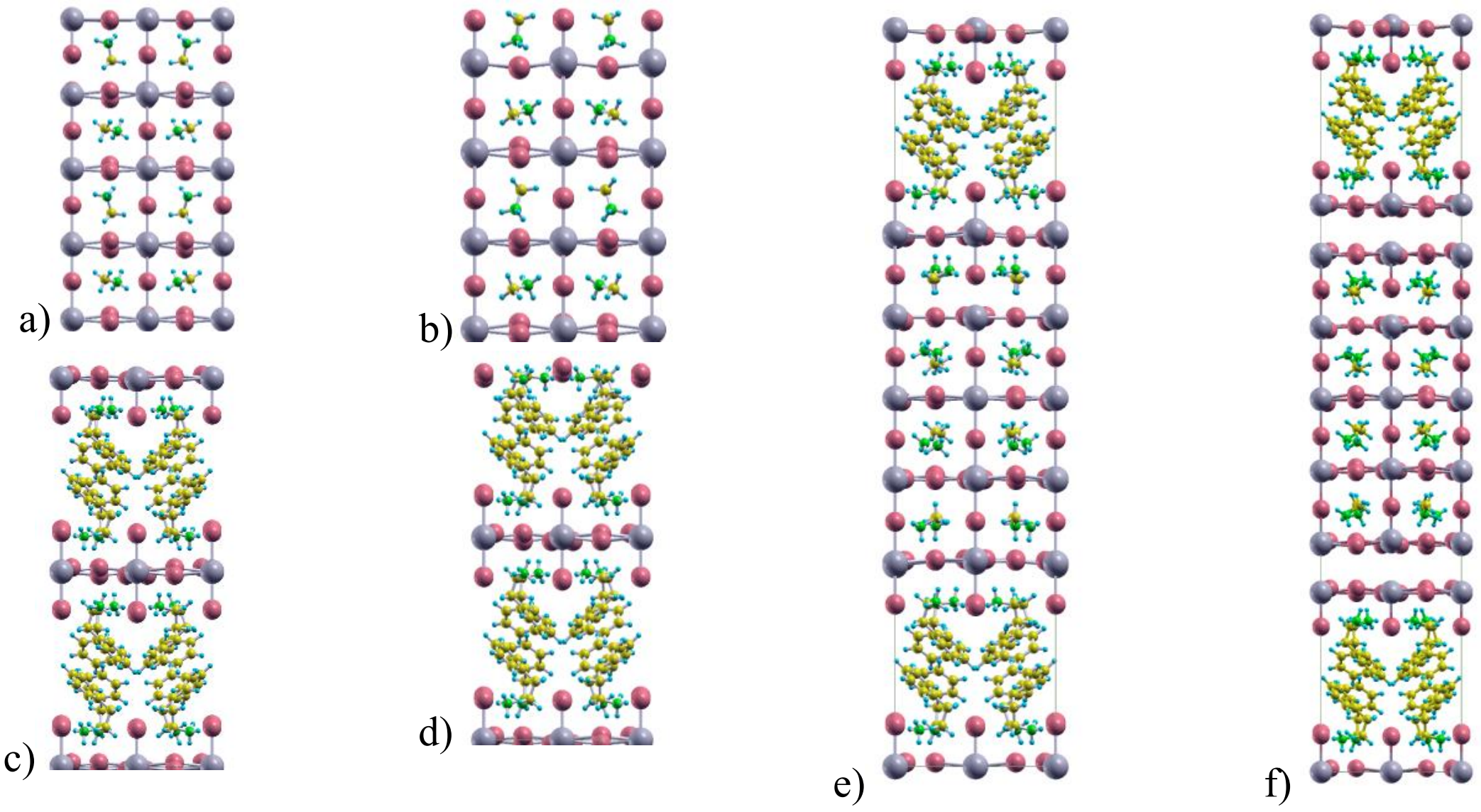

**Figure 3:** a) – d) [100] view of the top the (001) surface of $Cs_{0.125}MA_{0.14}FA_{0.735}Pb(I_{0.87}Br_{0.13})3$ (3D) and $(PEA)_2PbI_4$ (2D): a) and b) $PbI_2$ and (Cs/FA/MA)I termination of $Cs_{0.125}MA_{0.14}FA_{0.735}Pb(I_{0.87}Br_{0.13})_3$, respectively; c) and d) $PbI_2$ and PEAI terminations of $(PEA)_2PbI_4$, respectively. e) – f) [100] view of the models of the interfaces 3D/2D: e) 3D with $PbI_2$ terminations with 2D with PEAI termination and f) 3D with $PbI_2$ terminations with 2D with $PbI_2$ termination Pb in gray, I in red, C in yellow, N in green and H in blue.

The use of this approach for the determination of band alignments and band offsets at interfaces of the two complex 2D and 3D perovskites chosen for this work is an ongoing aspect of this study. This work always focuses on their (001) surfaces which can have two types of termination as shown in figure 3: for instance, the top of the 3D surfaces can be terminated by $PbI_2$ or (MA/FA/Cs)I moieties.

These various types of cleavage and terminations will influence the chemical reactivity of the surfaces, WF and EA, and the bands offset and alignment at the interfaces, as illustrated later. The interface model leads to expensive calculation because both surfaces have to be thick enough to have converged electronic properties, and the interfaces area must be large enough to compensate for the mismatch between materials.

Preliminary tests have been conducted on $MAPbI_3/(PEA)_2PbI_4$ interfaces. The chosen models involved MAPI (001) with $PbI_2$ termination interfaced with PEAPI (001) having either PEAI or $PbI_2$ terminations, see figure 4. These systems, comprising up to 600 atoms and fully relaxed, exhibited a small lattice mismatch (< 1 %).

The obtained band gaps for the full interface systems were 1.81 eV for the $MAPbI_3$-$PbI_2/(PEA)_2PbI_4$-PEAI interface and 2.03 eV for the $MAPbI_3$-$PbI_2/(PEA)_2PbI_4$-PbI2 interface, which are intermediate to the band gaps of the individual compounds. Preliminary values for the average electrostatic potential lineup ($\Delta V$) were 0.70 eV and 0.96 eV for the $MAPbI_3$-$PbI_2/(PEA)_2PbI_4$-PEAI and $MAPbI_3$-$PbI_2/(PEA)_2PbI_4$-$PbI_2$ interfaces, respectively. This difference is consistent with the $MAPbI_3$-$PbI_2/(PEA)_2PbI_4$-$PbI_2$ interface possessing a more insulating character.

Further investigations have focused on more complex 2D/3D perovskite interfaces. The preliminary results for the 2D/3D/2D interfaces reveal significant insights into band alignment, see figure 5. This figure gives the macroscopic potential through this interface. It clearly shows that the macroscopic potential across these interfaces is directly influenced by the chemical nature of the constituent compounds. A notable slope in the potential is observed, which is linked to the polarization of the 3D materials resulting from the substitution of FA by Cs and MA cations, and I by Br. At this stage, the contribution of each ion has not been evaluated but it is necessary to remind that MA and FA possess strong dipole moments (± 2 Debye) which can induce polarized dipole order in at the nano-micro scales in order to minimize the electrostatic dipole-dipole energy or interaction [21 – 23]. Furthermore, the band alignment is highly dependent on the specific termination and chemical nature of the perovskites at the interface. As illustrated by figure 6, the valence band offset can vary significantly, ranging from 0.2 eV to 0.7 eV, while the conduction band offset is approximately 0.25 eV (subject to further confirmation) without necessarily the band gap of the full system.

These findings underscore the critical role of interface engineering in controlling charge transport in such devices.

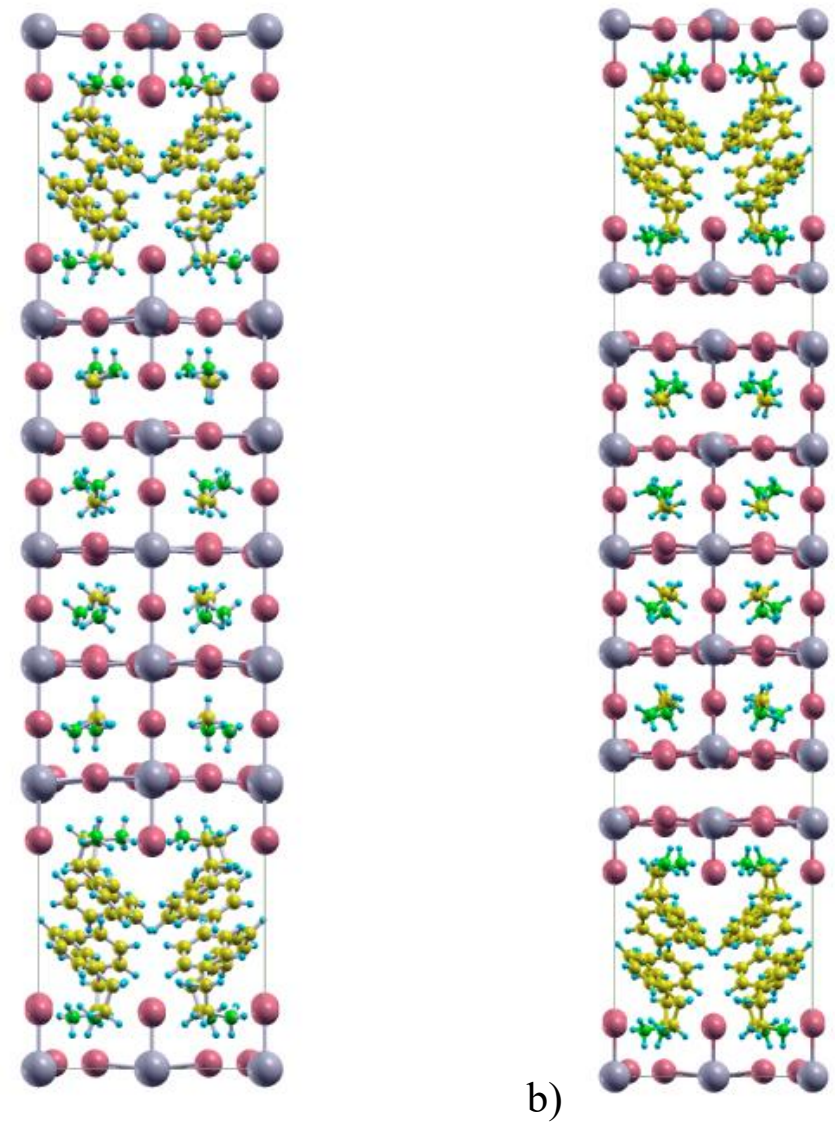

**Figure 4:** [100] view of the models of the interfaces $MAPbI_3/(PEA)_2PbI_4$: a) MAPI with $PbI_2$ termination with PEAPI with PEAI termination; b) MAPI with $PbI_2$ termination with PEAPI with $PbI_2$ termination. Pb in gray, I in red, C in yellow, N in green and H in blue.

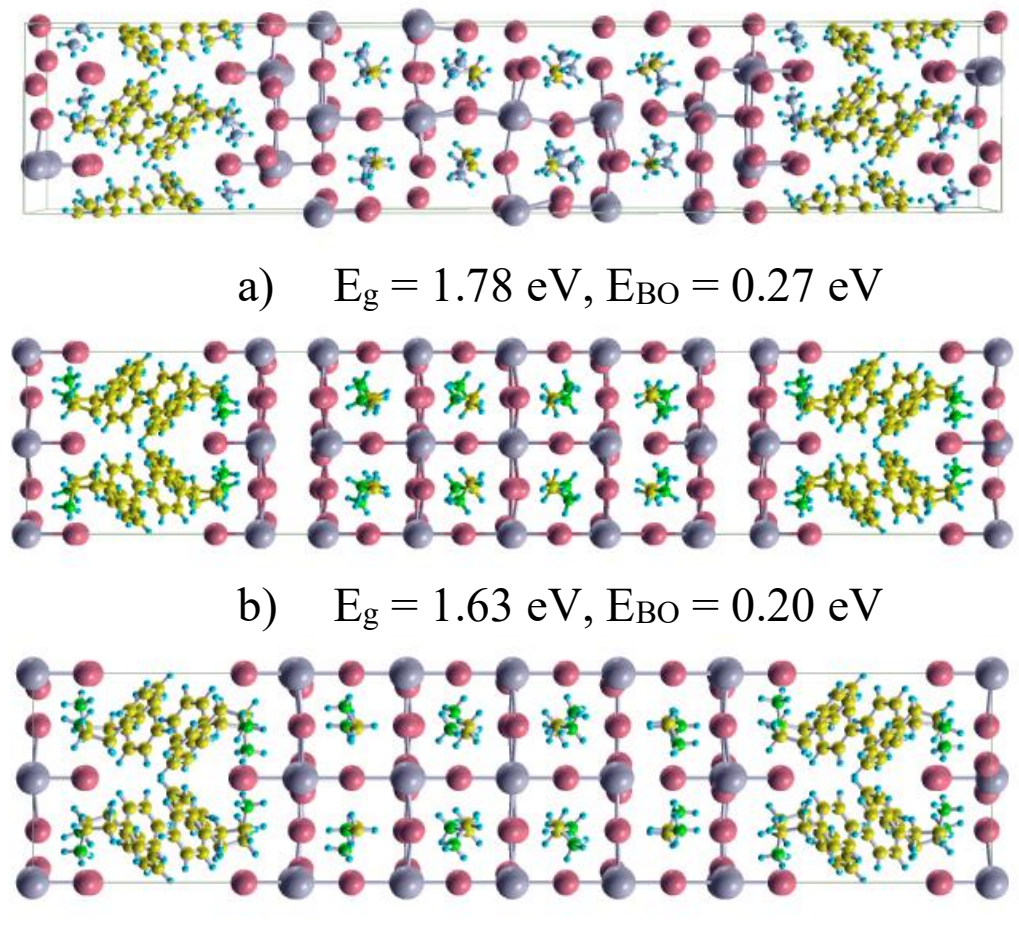

a) $E_g$ = 1.78 eV, $E_{BO}$ = 0.27 eV

b) $E_g$ = 1.63 eV, $E_{BO}$ = 0.20 eV

c) $E_g$ = 1.69 eV, $E_{BO}$ = 0.67 eV

**Figure 6:** Example of the impact on the band gaps and band offsets of the cleavage of the [100] $(Cs/FA/MA)Pb(Br/I)_3/(PEA)_2PbI_4$ interfaces: a) $(Cs/FA/MA)Pb(Br/I)_3$ with $PbI_2$ termination and $/(PEA)_2PbI_4$ with $PbI_2$ termination fully optimized; b) $(Cs/FA/MA)Pb(Br/I)_3$ with $PbI_2$ termination and $/(PEA)_2PbI_4$ with $PbI_2$ termination in phase; c) $(Cs/FA/MA)Pb(Br/I)_3$ with $PbI_2$ termination and $/(PEA)_2PbI_4$ with PEAI termination. The thickness of the used slab is 13 nm.

## 4 CONCLUSIONS

In this study, hybrid functional approaches based on first-principles calculations have been applied to investigate the structural and electronic properties of complex 2D/3D perovskites interfaces. As shown in previous work, the results for the valence band determination and electron affinities are consistent with available published data. The preliminary results demonstrate their dependence on the cleavage direction and chemical nature of the surfaces.

The work on band offset and alignment determination at interfaces is still in progress, but preliminary findings highlight the significant impact of interface creation on macroscopic potential and the resulting band offset and alignment. The chemical composition and termination of the perovskite layers at the interface are crucial determinants of the valence and conduction band offsets.
Future work will involve applying these methodologies to other relevant interfaces to further broaden our understanding of band alignment in tandem solar cell architectures. But given the complexity of these kinds of systems, the feasibility of this type of study depends heavily on access to significant HPC resources: To speed up the process, the use of other atomistic methods (such as the combination of the molecular dynamic and effective Hamiltonian [24]), and, the coupling with machine learning/IA approaches and targeted experimental characterizations would be highly valuable for identifying existing species and defects, thereby facilitating the understanding and the optimization of these intricate systems.

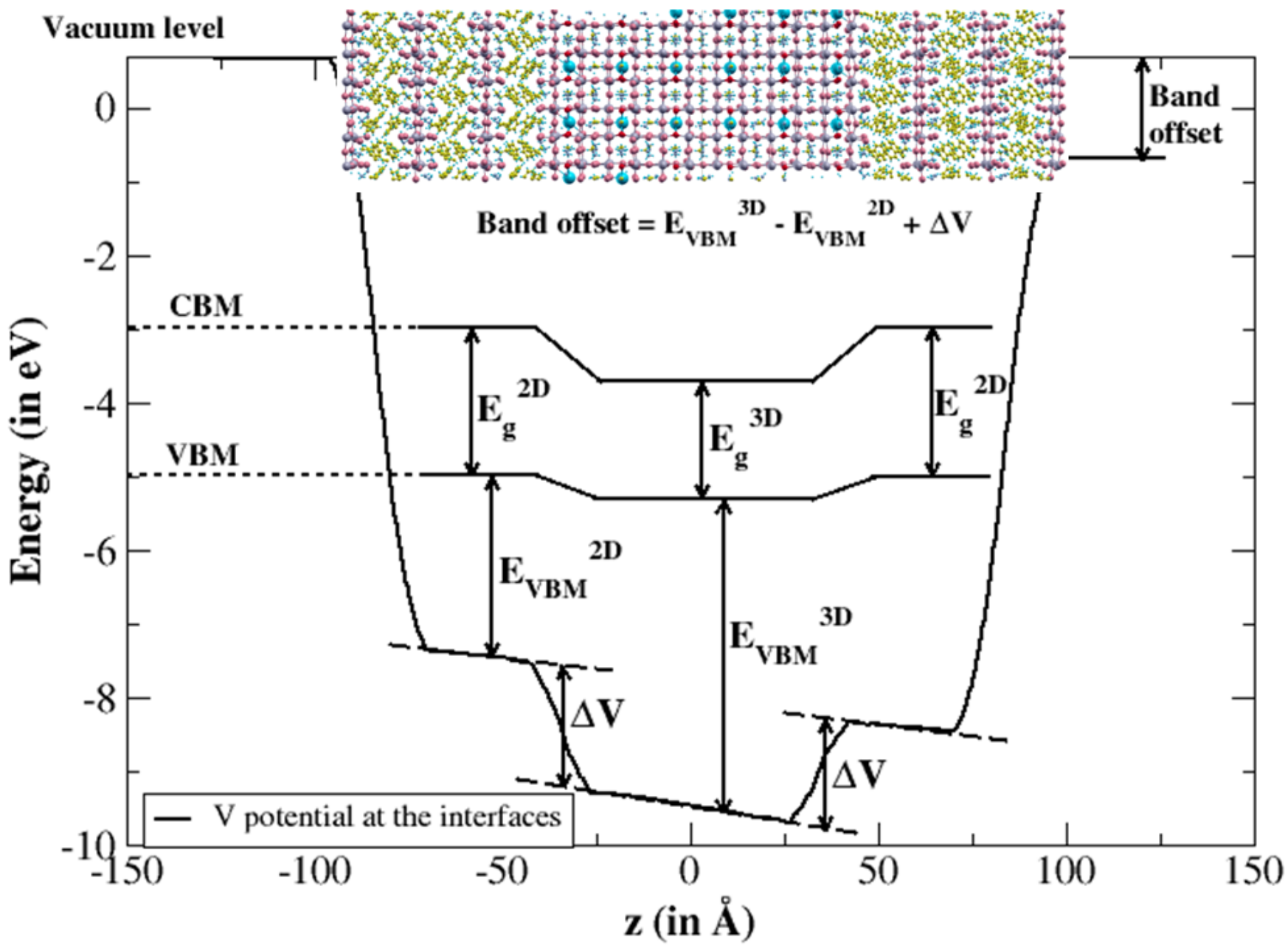

**Figure 5:** Band alignment through the 2D/3D/2D perovskites interface showing the band structure and macroscopic potential as a function of atomic position. The band structure of the isolated 2D and 3D surfaces are given for reference. For seek of clarity, only macroscopic potential is given to show the impact of the interface formation. The band structure at the interface can be obtained by the correction of the vacuum level.

Acknowledgments

The author thanks the support from the France 2030 program PEPR-TASE ("Programme et Equipements Prioritaires de Recherche sur les Technologies Avancées des Systèmes Energétiques") specifically within the MINOTAURE project Grant ANR-22-PETA-0015.

References

[1] M.A. Green *et al.* Prog. Photovolt. Res. Appl. 33 (2025) 795.
[2] https://www.nrel.gov/pv/cell-efficiency
[3] Z.G. Karabag *et al.*, Adv. Energy Mater. 13 (2023) 2302038.
[4] Ph. Baranek *et al.* (2025) https://doi.org/10.4229/EUPVSEC2025/2AO.3.5.
[5] Ph. Baranek *et al.*, EPJ Photovoltaics 17 (2026) 33.
[6] A. Gissler *et al.*, Solar RRL 10 (2026) e202500690.
[7] C.G. Van de Walle *et al.*, Phys. Rev. B 35 (1987) 8154.
[8] A. Baldereschi *et al.*, Phys. Rev. Lett. 61 (1988) 734.
[9] C. Sgiarovello *et al.*, Phys. Rev. B 64 (2001) 195305.
[10] I. Borriello *et al.*, Phys. Rev. B 76 (2007) 035430.
[11] T. Bischoff *et al.*, Phys. Rev. B 101, 235302 (2020).
[12] F. Lafond *et al.*, J. Phys. Chem. 124 (2020) 10353.
[13] R. Dovesi *et al.* WIREs Comput. Mol. Sci. 8 (2018) e1360.
[14] R. Dovesi *et al.* CRYSTAL23 User's Manual (University of Torino, Torino, 2023).
[15] https://www.crystal.unito.it
[16] J.P. Perdew *et al.* Phys. Rev. Lett. 100 (2008) 136406.
[17] D. Ory *et al.*, Adv. Opt. Mat. 12 (2024) 202401212.
[18] B. Traore *et al.*, Phys. Rev. Mat. 6, 014604 (2022).
[19] A. Mishra *et al.*, Surfaces and Interfaces 25 (2021) 101264.
[20] S. Mejaouri *et al.* Small Methods 8 (2024) 230091.
[21] J.M. Frost *et al.* Nano. Lett. 14 (2014) 2584.
[22] J.M. Frost *et al.* APL Mater. 2 (2014) 08506.
[23] J. Järvi *et al.* New J. Phys. 20 (2018) 103013.
[24] M. Schwade *et al.*, Nat. Commun. 17 (2026) 2652.